\documentclass[12pt]{article}
\let\large=\normalsize
\newcommand{\be}[3]{\begin{equation}  \label{#1#2#3}}     
\newcommand{\ee}{ \end{equation}}
\newcommand{\ba}{\begin{array}}
\newcommand{\ea}{\end{array}}

\newcommand{\PRL}[2]{{\em Phys. Rev. Lett.}{ \bf #1#2}}

\newcommand{\href}[1]{{\ref{#1}}}

\begin{document}
\thispagestyle{empty}
\rightline{HUB-EP-97/47}
\rightline{hep-th/9708103}
\vspace{2truecm}
\centerline{\bf GENERAL BPS BLACK HOLES IN FIVE DIMENSIONS}
\vspace{1.2truecm}
\centerline{\bf 
W. A. Sabra\footnote{sabra@qft2.physik.hu-berlin.de}}
\vspace{.5truecm}
{\em 
\centerline{Humboldt-Universit\"at, Institut f\"ur Physik}
\centerline{Invalidenstra\ss e 110, 10115 Berlin, Germany}
}
\vspace{1.2truecm}
\vspace{.5truecm}
\begin{abstract}
We construct general static black hole configuration for the theory of
$N=2$, $d=5$ supergravity coupled to an arbitrary number of Abelian 
vector multiplets. The underlying $very$ $special$ $geometry$ structure plays 
a major role in this construction. From the viewpoint of M-theory compactified
on a Calabi-Yau threefold, these black holes are identified with BPS winding
states of the membrane around 2-cycles of the Calabi-Yau threefold, and thus
are of importance in the probing of the phase 
transitions in the moduli space of M-theory 
compactified on a Calabi-Yau threefold.
\end{abstract}
\bigskip
\bigskip
\newpage

\section{Introduction} 
Recently progress has been made in the
study of extreme solutions of $N=2$ supergravity in four
and five dimensions. These solutions have a rich structure in 
contrast to those of $N=4, 8$
supergravity theories, where the dynamics is severely restricted by 
supersymmetry. Namely, black holes of $N=4, 8$ supersymmetric theories 
do not receive quantum corrections whereas solutions of $N=2$ supergravity
alter significantly when quantum corrections are taken into
account. The starting point in the understanding of non-singular static 
$N=2$ black 
holes came with the realization that their Bekenstein-Hawking entropy, 
${\bf S}_{BH},$ 
\cite{bh} can be obtained by extremizing the central charge 
of the underlying $N=2$ super-algebra 
with respect to the moduli \cite{fe/ka1},\cite{fe/ka2},\cite{re}. 
The entropy of these black holes was found to
satisfy the following universal formula
\begin{equation}
{\bf S}_{BH}\sim\vert Z_{fix}\vert^{\alpha}
\end{equation}
where $\alpha=2,{3\over2}$ for $d=4,5$ respectively and 
$Z_{fix}$ is the extremal value of the central charge. 

Moreover, it was shown (for the 4-dimensional case), 
that the scalar fields follow attractor equations with 
fixed points on the horizon \cite{fe/ka1}. These fixed values  
are independent of the values of the scalar  fields at infinity 
(moduli) and are determined only in terms of conserved charges and 
topological data. 
If one assumes that the scalar fields take these fixed values 
throughout the entire space-time, one obtains in this way the 
so-called double extreme black holes \cite{ka/sh}.

The above results are covariantly formulated in terms of the underlying 
special geometry \cite{special}. For instance, the equations 
which give the values of the scalar fields near the horizon,
the so-called stabilisation equations for the double-extreme black
holes are  given by \cite{ka/sh}
\begin{equation}
i (\bar ZL^I -Z\bar L^I ) = p^I \qquad , \qquad 
i (\bar ZM_I -Z\bar M_I ) = q_I,
\label{sta}
\end{equation}
where $(L^I,M_I)$ are the covariantly holomorphic section 
of $N=2$ supergravity \cite{special}, 
$p^I$ and $q_I$ are, respectively, integer valued 
magnetic and electric charges. 
By means of the equations (\ref{sta}), one can, in principle  
compute the classical entropy of static $N=2$ 
black holes \cite{be/ka1}, \cite{sa1} as well as to incorporate quantum
corrections \cite{be/ca}, \cite{be/ga}.\footnote{The quantum corrections are
encoded in the $M_I$ part of the covariantly holomorphic section}. 
For example, in type II string compactifications,
the entropy depends on the topological data of the Calabi-Yau 
spaces, like the intersection numbers, the Euler number and the rational 
worldsheet instanton numbers \cite{be/ca}.

Later extreme solutions, with non-constant values for the moduli,
were derived, by solving for the equations of motion, 
 for the axion-free $STU$ model with cubic
prepotential, and for supergravity models based on quadratic 
prepotentials \cite{Klaus},\cite{sa1}. 
It was then shown in \cite{sa2}
that for general static extreme
$N=2$ black holes, the solutions are completely specified
by the K\"ahler potential of the underlying special geometry where the 
imaginary part of the holomorphic sections are given in terms of a 
set of constrained harmonic functions. These constraints correspond to
asymptotic flatness and the vanishing of the K\"ahler connection. 
Moreover, it was shown in \cite{bls} that the metric for stationary, 
but non-static solutions depends also on the
$U(1)$ K\"ahler connection.
Depending on the choices for the harmonic functions and on the
considered prepotentials, one gets for example non-static
rotating $N=2$ black holes,  $N=2$ Taub-NUT spaces or $N=2$ 
Eguchi-Hanson like instantons.

More concretely, in four dimensions, 
it is very well known from the work of Tod \cite{to}, 
that the most general form of the metric admitting supersymmetries can
be written in the IWP \cite{is/wi} metric form  
\begin{equation}
ds^2 = -e^{2U}(dt + \omega_m dx^m)^2 +e^{-2U}dx^m dx^m,
\label{gg}
\end{equation}
and $\omega_m$ is defined by 
\begin{equation}
\vec\nabla\times \vec\omega=-{i\over (V\bar V)^2}(\bar V\vec\nabla
V-V\vec\nabla\bar V)
\end{equation}
where $e^{2U} =V\bar V$ and $V$ is the inverse of a harmonic function.

For $N=2$ supergravity models with vector multiplets, 
the metric is of the form (\ref{gg}), 
with 
\be170
\ba{rcccl}
e^{-2U}&=&e^{-K} & \equiv & i (\bar{X}^I F_I- X^I \bar{F}_I) \ , \\
{1 \over 2} e^{2U} \epsilon_{mnp} \partial_n \omega_p&=&Q_m & \equiv &
 {1 \over 2} e^{K}( \bar{F}_I \partial_m X^I - \bar{X}^I \partial_m F_I
 + c.c. ) \\
&&&=& { 1 \over 2} e^{2U} ( H_I \partial_m \tilde{H}^I - \tilde{H}^I
\partial_m H_I) 
\ea
\ee
\
where the equations which define the moduli fields as well 
the electric and magnetic fields strengths are given by 
\begin{eqnarray}
i (X^I - \bar{X}^I ) &=& \tilde{H}^I(x^{\mu}) \qquad , \qquad 
i (F_I -\bar{F}_I) = H_I(x^{\mu})\label{gse}\\ 
F^I_{mn} &=& {1 \over 2} \epsilon_{mnp} \partial_p \tilde{H}^I
\qquad , \qquad G_{I\, mn} = {1 \over 2} \epsilon_{mnp} \partial_p H_I \
\end{eqnarray}
here $(X^I , F_I)$ is the symplectic holomorphic section, and 
$(H_I, \tilde H^I)$ are harmonic functions.
For vanishing K\"ahler connection ($Q_m=0$), one obtains static solutions.

Using the explicit static black hole solutions, one can calculate the entropy
and the value of the scalar fields near the horizon. It can be easily
demonstrated that the entropy is given in terms of the extremum of the central
charge and that the equations (\ref{gse}) reduce to those of (\ref{sta}) near
the horizon \cite{sa2}. 

In five dimensions, supersymmetric black holes were constructed for the case
of pure $N=2$ supergravity in \cite{gkltt}. The metric in this case 
is of the Tanghelini form \cite{frt},\cite{perry}. 
In the context of supergravity with vector multiplets, double-extreme black
holes were constructed and their entropy was calculated in terms of the
extremised central charge in \cite{chams}. Moreover, in \cite{chams},
the Strominger-Vafa black hole \cite {SV} was
reproduced as a double-extreme black hole of an $N=2$ supergravity model with
one vector multiplet. Also the rotating black hole solution of \cite{ppp} 
was embedded into an N=2 supergravity model with one vector multiplet, 
and was shown to admit unbroken supersymmetry \cite{raja}.
Five dimensional static and rotating black holes were
also constructed in 
\cite{Tsey} and 
\cite{mirjam} in the context of heterotic and type II string toroidal 
compactifications.

In spite of lots of work on the five dimensional black holes, general 
supersymmetric solutions with non-constant values of the moduli, surprisingly, 
have not been constructed yet.
It is our purpose in this paper to find explicit black hole 
solutions of $N=2$ $d=5$ supergravity coupled to an arbitrary number of 
vector multiplets. Here we will only concentrate on the electrically charged 
static solution and leave the rotating solutions as well as the magnetic black
string solutions for a separate publication.
In five dimensions, the couplings of  
$N=2$ supergravity to Abelian vector 
multiplet, is based on the structure of very special geometry and thus one
expects that this geometric structure should play a major role in the study
of black hole solutions similar to that played by 
special geometry in four dimensions.

This work is organised as follows.
In the next section,  we briefly review the structure of $N=2$ supergravity
in $d=5$ and collect some formulae and expressions
which will be relevant for our later discussion.
In section three,  we will give the static black hole solutions and 
verify that they admit
unbroken supersymmetry by solving for the
supersymmetry transformation rules for the gravitino and the gauginos in a
bosonic background. We will also study the behaviour of our solutions near the
horizon and demonstrate that the values of the moduli take fixed values which
extremise the central charge. We also derive the expression of the
entropy in terms of the extremised central charge, and give the general form
of the double-extreme black hole solutions. The last section contains a
summary of our results and a discussion on the relevance of the black hole
solutions to the analysis of topological phase transitions of M-theory on a
Calabi-Yau 3-folds.
\section {$d=5$ $N=2$ Supergravity and Very Special Geometry}
The theory of $N=2$ supergravity theory coupled to an arbitrary number $n$ of
Maxwell's supermultiplets was first considered in \cite{GST}. 
Recent discussions of the bosonic part 
(our main concern in this work) in terms of very 
special geometry is given in \cite{vpdw}. Also, the construction of $N=2$
supergravity as a compactification of  $d=11$ supergravity 
down to five dimensions on Calabi-Yau 3-folds
was discussed  more recently in \cite{Cadavid}. 

In the original analysis of \cite{GST}, it was established that 
the scalar fields of the vector multiplets parametrise a riemannian space.
Those which are homogeneous symmetric spaces take the form,
\begin{equation}
{\cal M}={\hbox{Str}_0(J)\over \hbox{Aut}(J)},
\end{equation}
where $\hbox{Str}_0(J)$ is the reduced structure group of a 
formally real unital Jordan Algebra of degree three, 
$\hbox{Aut}(J)$ is its automorphism group. 
The classification 
of homogeneous symmetric spaces was then reduced to that of Jordan algebras of
degree three. Moreover, the scalar manifold can be regarded as a hyperspace,
with vanishing second fundamental form of an $(n+1)$-dimensional riemannian
space $\cal G$ whose coordinates $X$ are in correspondence with the vector
multiplets including that of the graviphoton. The equation of 
the hypersurface is ${\cal V}=1$, where 
$\cal V$ , the prepotential, is a homogeneous cubic polynomial in the 
coordinates of $\cal G,$
\begin{equation}
{\cal V} = {1\over 6} C_{IJK} X^I X^J X^K \ .
\label{pin}
\end{equation}

Non-simple Jordan algebras of degree three are of the form $R\oplus\Sigma_n$,
where $\Sigma_n$ is the Jordan algebra associated with a quadratic form. The
corresponding symmetric scalar manifold are the form 
\begin{equation}
{\cal M}={SO(1,1)}\times {SO(n-1,1)\over SO(n-1)}.
\label{man}
\end{equation}
In this case, $\cal V$ is factorisable into a linear times a quadratic form
in $(n-1)$ scalars, which for the positivity of the kinetic terms 
in the Lagrangian, must have a Minkowski metric.
For simple Jordan algebras, one obtains four sporadic locally symmetric spaces
related to the four simple unital formally real Jordan algebras over the four
division algebras of real, complex, quaternions and octanions. For more
details we refer the reader to \cite{GST}.

For M-theory compactifications on a Calabi-Yau threefold with Hodge numbers
$h_{(1,1)}$, $h_{(2,1)},$ and intersection numbers $C_{IJK}$, 
$(I, J, K=1,\cdots h_{(1,1)}),$ ${\cal V}$ represents
the intersection form of the Calabi-Yau threefold related to the overall 
volume of the Calabi-Yau threefold and belongs to the so-called 
universal hypermultiplet.
Also, in this picture, $X^I$ correspond to the size of the 2-cycles of the
Calabi-Yau threefold. The massless spectrum of the theory contains
$h_{(1,1)}-1$ vector multiplets with real scalar components defined by the
moduli at unit volume. Including the graviphoton, the theory has $h_{(1,1)}$
vector bosons. In addition to the universal hypermultiplet present in any
Calabi-Yau compactification, the theory also
contains $h_{(2,1)}$ hypermultiplets.\footnote{Briefly, in the 
11-dimensional theory, we have the 
metric $G_{\hat\mu\hat\nu}$ and a three-form gauge field 
${\cal A}_{\hat\mu\hat\nu\hat\rho}$. If we split
$\hat\mu$ into $(\mu, a,\bar a)$, with $(\mu=1,\cdots,5)$ the space-time
indices and $(a,\bar a)$
represent the three internal complex dimensions of the Calabi-Yau, then the
$h_{(1,1)}-1$ scalar moduli correspond to $G_{a\bar b}$ defined at 
unit volume ($det G_{a\bar b}=1.$) The vector bosons correspond to the
$h_{(1,1)}$ space-time one-forms ${\cal A}_{\mu a \bar a}$. The universal
hypermultiplet contains $({\cal V}, {\cal A}_{\mu\nu\rho}, {\cal A}_{abc}=
\epsilon_{abc}C)$, 
the $h_{(2,1)}$ hypermultiplets correspond to 
$(G_{ab},{\cal A}_{ ab \bar c})$}

We now turn to discuss the effective 
$N=2$ supersymmetric Lagrangian describing the 
coupling of vector multiplets to supergravity in five dimensions. This
Lagrangian is determined entirely in terms of a 
cubic prepotential (\ref{pin}), which defines very special 
geometry \cite{vpdw}.
For our purpose, we only need to display the bosonic part of 
the Lagrangian and the 
supersymmetry transformation
laws for the gravitino and the gauginos. 
The bosonic action is
\begin{equation}
e^{-1} {\cal L} = -{1\over 2} R - {1\over 4} G_{IJ} F_{\mu\nu} {}^I
F^{\mu\nu J}-{1\over 2} g_{ij} \partial_{\mu} \phi^i \partial^\mu \phi^j
+{e^{-1}\over 48} \epsilon^{\mu\nu\rho\sigma\lambda} C_{IJK} 
F_{\mu\nu}^IF_{\rho\sigma}^JA_\lambda^k
\end{equation}
where $R$ is the scalar curvature, $F_{\mu\nu}=2\partial_{[\mu}A_{\nu]}$ 
is the Maxwell field-strength
tensor and $e=\sqrt{-g}$ is the determinant of the F\"unfbein 
$e_\mu^{\underline \nu}$. 
\footnote {In this paper, we will use for the metric the signature $(-++++)$ 
and for the indices we take: $\mu, \nu .. = 0,1,2,3,4$, whereas 
$m,n,\cdots = 1,2,3,4$.  We 
denote the underlined indices
$\underline{\mu} , \underline{\nu}$ as flat ones.
Antisymmetrized indices are defined by: $[\mu\nu]
= {1 \over2} (\mu\nu -\nu\mu)$.}

The fields $X^I= X^I(\phi)$ are the
special coordinates satisfying
\begin{equation}
X^I X_I=1 , \qquad {1\over 6}C_{IJK} X^I X^J X^K =1
\label{cn}
\end{equation}
where, $X_I$, the dual coordinate is defined by,
\begin{equation}
X_I={1\over6}C_{IJK}X^JX^K.
\label{d}
\end{equation}

The moduli- dependent gauge coupling metric is related to the prepotential
(\ref{pin}) via
the relation

\begin{equation}
G_{IJ} = -{1\over 2}{\partial\over \partial X^I}
{\partial\over\partial X^J}(\ln {\cal V})|_{{\cal V} =1}
\label{gc}
\end{equation}
The metric $g_{ij}$ is given by
\begin{equation}
g_{ij}=  G_{IJ} \partial_{i}X^I\partial_{j}X^J|_{{\cal V} =1};\qquad  
(\partial_i \equiv {\partial \over \partial\phi^i})
\label{metric}
\end{equation}

The supersymmetry transformation laws for the Fermi fields in a 
bosonic background are given by \cite{GST}
\begin{eqnarray}
\delta\psi_\mu &=& {\cal D}_\mu\epsilon + { i\over 8}
X_I
\Bigl(\Gamma_\mu{}^{\nu\rho} - 4 \delta_\mu{}^ \nu \Gamma^\rho\Bigr)
F_{\nu\rho}{}^I \epsilon,\nonumber\\
\delta \lambda _i &=&{3\over 8}\partial_iX_I\Gamma^{\mu\nu}\epsilon
F_{\mu\nu}^I
- {i\over 2} g_{ij} \Gamma^\mu \partial_\mu\phi^j \epsilon ,\nonumber \\.
\label{stl}
\end{eqnarray}
where 
$\epsilon$ is the supersymmetry parameter and 
\begin{equation}
{\cal D}_\mu=\partial_\mu+{1\over4}\omega_{\mu\underline{\nu}\underline{\rho}}
\Gamma^{\underline{\nu}}\Gamma^{\underline{\rho}}
\end{equation}
Here, $\omega_{\mu\underline{\nu}\underline{\rho}}$ is the spin connection, 
$\Gamma^{{\nu}}$ are Dirac matrices and 
\begin{equation}
\Gamma^{{\nu_1}{\nu_2}\cdots{\nu_n}}=
{1\over n!}\Gamma^{[{\nu_1}}\Gamma^{{\nu_2}}
\cdots \Gamma^{{\nu_n}]}.
\end{equation}

\section{Extreme Black Hole Solutions}
In this paper, we are interested in electrically charged BPS black hole 
solutions, $i. e$ configurations that break half of the supersymmetry. 
In the context of
M-theory compactification on a Calabi-Yau, these correspond to BPS 
winding states of the membrane around
2-cycles of the Calabi-Yau threefold. Magnetically charged (string)
solutions, which in the M-theory picture correspond to winding BPS states of
the five-brane around the 4-cycles,
together with general stationary electrically charged solutions 
will be discussed elsewhere \cite{wt}.

Our ansatz for the BPS black hole solution is 
\begin{eqnarray}
ds^2 &=&-e^{-4 U} dt^2 +e^{2 U} (d\vec{x})^2\nonumber\\
G_{IJ} F_{0m}^I&=&{1\over2}e^{-4U}\partial_mH_J, \qquad 
{\tilde X_I}=e^{2U}X_I={1\over 3}H_I\nonumber\\
e^{3U}&=&{1\over 6}C_{IJK}{\tilde X}^I{\tilde X}^J{\tilde X}^K={\tilde
X_I}{\tilde X^I},\qquad \tilde X^I=e^{U}X^I\nonumber\\
H_I&=&h_I+{q_I\over r^2}
\label{magic}
\end{eqnarray}
where $h_I$ are constants related to the values of the scalars at infinity and
$q_I$ are electric charges.

Before we proceed to show that the above configuration is indeed a BPS one, 
we will derive some relations which will be useful for our purposes.
First if one differentiate (\ref{cn}) with respect to the scalar fields, 
keeping in mind that $C_{IJK}$ is a 
constant symmetric tensor we obtain
\begin{equation}
\partial_iX_{I} = {1\over3}C_{IJK} \partial_iX^J X^K\, \qquad
X^I\partial_iX_I=X_I\partial_iX^I=0.
\label{si}
\end{equation}
Moreover, the gauge coupling metric $G_{IJ}$ can be expressed in terms of the
special coordinates by
\begin{equation}
G_{IJ}=-{1\over2} C_{IJK}X^K+{9\over2}X_I X_J
\label{gcs}
\end{equation}
In order to relate the derivative of the special coordinate to that of 
its dual, we
use (\ref{gcs}) together with (\ref {si}), this gives the following expression
\begin{equation}
\partial_i X_I=-{2\over3}G_{IJ}\partial_iX^J.
\label{scg}
\end{equation}
Finally it is very useful to find an expression for the graviphoton
field-strength ($X_IF^I$), in our Ansatz. In order to do this, 
we note that from (\ref{magic}) we obtain 
\begin{equation}
e^{2U}={1\over 3}H_IX^I.
\end{equation}
Differentiating both sides of the above equation and using the second equation
on (\ref{si}) yields
\begin{equation}
\partial_m(e^{2U})={1\over3}\partial_m(H_I)X^I
\end{equation}
Moreover, using (\ref {gcs}) and (\ref {d}) we obtain the following relation
\begin{equation}
X_I={2\over3}G_{IJ} X^J,
\end{equation}
therefore,
\begin{equation}
X_IF^{I}_{0m}={2\over3}
G_{IJ}X^JF^I_{0m}={1\over3}e^{-4U}X^J\partial_m(H_I)=-\partial_m(e^{-2U}).
\label{important}\end{equation}

The F\"unfbeins for the metric in (\ref{magic}) are
\begin{eqnarray}
e_{{0}}^{\ \underline 0} &=& e^{-2U} \quad , \quad 
e_{{m}}^{\ \underline n} = e^{U} \delta_{{m}}^{\ n}\nonumber \\
e_{{0}}^{\ \underline m} &=&  e_{{m}}^{\ \underline 0} = 0\nonumber\\ 
e^{{0}}_{\ \underline 0} &=& e^{2U} \quad , \quad 
e^{{m}}_{\ \underline n} = e^{-U} \delta^{{m}}_{\ n}\nonumber\\
e^{{0}}_{\ \underline m} &=&  e^{{m}}_{\ \underline 0} = 0
\label{funf}
\end{eqnarray}

For the spin connections one obtains
\footnote{\ $\omega_{\mu}^{ab} = 2 e^{\nu[a} \partial_{[\mu} e_{\nu]}^{\ b]}
 - e^{\rho a} e^{\sigma b} e_{\mu p } \partial_{[\rho} e_{\sigma]}^{\ p}$}
\begin{equation}
\omega_{0}^{\underline 0\underline m} = e^{-U}\partial_m(e^{-2U}),\qquad
\omega_{m}^{\underline n\underline p} = \delta_m^{\ n}\partial_pU-
\delta_m^{\ p}\partial_nU
\label{sc}\end{equation}
with the rest vanishing.

Now we proceed to show that the configuration defined 
by (\ref{magic}) is supersymmetric.
First we start with the time component of the gravitino variation,
\begin{equation}
\delta\psi_0 = D_0 \epsilon +  { i\over 8}
X_I\Gamma_0{}^{\nu\rho}F_{\mu\rho}{}^I - {i\over2}
\Gamma^\rho X_IF_{0\rho}{}^I\epsilon,
\end{equation}
since the only non-vanishing component of the field strength is $F_{0m}$, the
second term on the right hand side vanishes and the above equation becomes
\begin{equation}
\delta\psi_0 ={1\over2}\omega_{0}^{\underline 0\underline m}
\Gamma^{\underline m}
\Gamma^{\underline 0}\epsilon - {i\over2}
\Gamma^\rho X_IF_{0m}{}^I\epsilon,
\end{equation}
then using (\ref{sc}) and (\ref{important}) as well as the relation
\begin{equation}
\Gamma^m=e^{-U}\Gamma^{\underline m}
\end{equation}
we obtain, 
\begin{equation}
\delta\psi_0 = 0 \Longleftrightarrow \Gamma^{\underline 0}\epsilon=-i\epsilon
\label{se}
\end{equation}
As a consequence generically one half of supersymmetry is broken.

Next, we turn to the space component for the gravitino supersymmetry
transformation law, this is given by
\begin{equation}
\delta\psi_m= D_m\epsilon + { i\over 8}X_I\Gamma_m{}^{\nu\rho}F_{\nu\rho}{}^I
\epsilon -{i\over2}\Gamma^0 X_IF_{0m}{}^I\epsilon,
\label{stg}
\end{equation}
where 
\begin{eqnarray}
{\cal D}_m\epsilon&=&\partial_m\epsilon+{1\over4}
\omega_{m}^{\underline n\underline p}
\Gamma^{\underline n}\Gamma^{\underline p}\nonumber\\
&=&\partial_m\epsilon+{1\over4}\partial_p U
[\Gamma^{\underline m}, \Gamma^{\underline p}]\epsilon\nonumber\\
{ i\over 8}X_I\Gamma_m{}^{\nu p}F_{\nu p}{}^I
\epsilon &=&-{i\over4}e^{2U}\partial_p(e^{-2U})\Gamma^{m0p}\epsilon={1\over4}
\partial_p U[\Gamma^{\underline p}, \Gamma^{\underline m}]\epsilon\nonumber\\
{ i\over 2}X_I\Gamma^0F_{0m}{}^I
\epsilon &=&\partial_mU\epsilon
\label{last}
\end{eqnarray}
where in deriving the above relations,
we have used the equation satisfied by the spinor (\ref{se}), our Ansatz for
the gauge fields and
\begin{equation}
\Gamma^{m0p}={1\over 2}\Gamma^0
[\Gamma^{p}, \Gamma^{m}]
={1\over2}\Gamma^{\underline 0}[\Gamma^{\underline p}, 
\Gamma^{\underline m}].
\end{equation}
Substituting the above expressions into (\ref{stg}), one obtains a simple
differential 
equation for the spinor $\epsilon$
\begin{equation}
(\partial_m+\partial_mU)\epsilon=0
\end{equation}
This gives
\begin{equation}
\epsilon=e^{-U}\epsilon_0
\end{equation}
where $\epsilon_0$ is a constant spinor satisfying 
$\Gamma^{\underline 0}\epsilon_0=-i\epsilon_0.$

We now turn to the gauginos supersymmetry variation given in (\ref{stl}). 
Using equations (\ref{metric}) and (\ref {scg}) this can be rewritten in the
form
\begin{eqnarray}
\delta \lambda _i &=&-{1\over 4}G_{IJ}\partial_iX^I\Gamma^{\mu\nu}\epsilon 
F_{\mu\nu}^I+{3i\over 4}\Gamma^\mu \partial_\mu X_I\partial_iX^I 
\epsilon\nonumber\\
&=&{1\over 2}G_{IJ}\Gamma^{m}\Gamma^{0}\partial_iX^I
F_{0m}^J\epsilon +{3i\over 4}\Gamma^m\partial_m X_I\partial_iX^I\epsilon
\end{eqnarray}
Using 
\begin{eqnarray}
G_{IJ}F_{0m}^J&=&{1\over2}e^{-4U}\partial_m(H_I),\qquad 
\partial_m X_I\partial_i X^I={1\over3}e^{-2U}\partial_m (H_I)\partial_i X^I
\nonumber\\
\Gamma^0&=&e^{2U}\Gamma^{\underline 0},\qquad
\Gamma^{\underline 0}\epsilon=-i\epsilon,\nonumber
\end{eqnarray}
it can be easily seen that the gaugino supersymmetry variation vanishes 
identically.
Thus, we have shown that the configuration (\ref{magic})
defines a supersymmetric bosonic configuration, which breaks
generically one half of $N=2$ supersymmetry. We note here that supersymmetry
does not restrict the functions $H_I.$ The Bianchi identities and equations of
motion for the gauge fields restrict $H_I$ to be harmonic functions.

Let us now look at the behaviour of our solution near the horizon. First we
write our solution in polar coordinates, this gives
\begin{equation}
ds^2 =-e^{-4 U} dt^2 +e^{2 U}(dr^2+r^2d\Omega_3^2).
\end{equation}
Near the horizon ($r\rightarrow 0$), $e^{2U}$ can be approximated as 
\begin{equation}
(e^{2U})_{hor}={1\over 3}(X^JH_J)_{hor}={1\over 3}(X^J)_{hor}({q_J\over r^2})=
{1\over 3}
{Z_{hor}\over r^2}
\label{horizon}\end{equation}
where $Z=q_IX^I$ is the central charge, 
and $Z_{hor}$ is its value at the horizon. 
The Bekenstein-Hawking entropy, ${\bf S}_{BH}$ related to the area of the 
horizon $\bf A$ is given by
\begin{equation}
{\bf S}_{BH}={{\bf A}\over 4G_N};\qquad 
{\bf A}=2\pi^2(r^2e^{2U})_{r=0}^{3\over2}=
2\pi^2\Big({Z_{hor}\over3}\Big)^{3\over2}
\end{equation}
where $G_N$ is Newton's constant.

Finally, using (\ref{horizon}), the second equation in (\ref{magic}) which 
defines the moduli over space time, becomes near the horizon
\begin{equation}
(ZX_I)_{hor}=q_I,
\end{equation}
which is the equation obtained from the extremization of the central charge 
\cite{chams},\cite{chaka}. If one assumes that the values of 
the moduli at the horizon are
valid throughout the entire space-time, then one obtains the double-extreme
black hole solution \cite{chams} 
where the metric takes the Tanghelini form \cite{frt},\cite{perry}
\begin{equation}
ds^2=-\Big(1+{Z\over 3r^2}\Big)^{-2}dt^2+
\Big(1+{Z\over 3r^2}\Big)(dr^2+r^2d\Omega_3^2)
.\end{equation}
\section{Example: STU=1}
In this section we discuss the black hole solution for the so-called $STU$
model. This model can be obtained from the compactification of heterotic 
string theory on $K_3\times S_1$ \cite{anton}. 
The tree-level prepotential in this model is 
given by $STU=1$ and corresponds to the scalar manifold of (\ref{man}) 
(for $n=2$). For this case we get the following equations  
\begin{eqnarray}
e^{2U}TU&=&H_0,\nonumber\\
e^{2U}SU&=&H_1,\nonumber\\
e^{2U}ST&=&H_2,
\label{skin}
\end{eqnarray}
where $H_0$, $H_1$ and $H_2$ are harmonic functions.
>From (\ref{skin}) we obtain the following solution for the metric and the
moduli fields,
\begin{equation}
e^{2U}=(H_0H_1H_2)^{{1\over3}}
\end{equation}
and 
\begin{eqnarray}
S&=&\Big({H_1H_2\over H_0^{2}}\Big)^{{1\over3}},\nonumber\\
T&=&\Big({H_0H_2\over H_1^{2}}\Big)^{{1\over3}},\nonumber\\
U&=&\Big({H_0H_1\over H_2^{2}}\Big)^{{1\over3}}.
\label{pain}
\end{eqnarray}
The metric solution is given by 
\begin{eqnarray}
ds^2&=&-(H_0H_1H_2)^{-{2\over3}}dt^2+(H_0H_1H_2)^{1\over3}
(dr^2+r^2d\Omega_3^2)\nonumber\\
\label{lina}
\end{eqnarray}
If the harmonic functions are represented by
$H_0={h_0+{q_0\over r^2}}$, $H_1={h_1+{q_1\over r^2}}$ and 
$H_2={h_2+{q_2\over r^2}}$,
then for an asymptotically flat metric one should demand that 
$h_0h_1h_2=1$
The ADM mass, $M_{ADM}$, of the black hole in five dimensions is given
by\footnote{see for example \cite{perry}}
\begin{equation}
g_{tt}=1-{8G_NM_{ADM}\over 3\pi}+\cdots
\end{equation}
Thus for our black hole,
\begin{equation}
M_{ADM}={\pi\over 4G_N}({q_0\over h_0}+{q_1\over h_1}+{q_2\over h_2})
\end{equation}
Eq. (\ref{pain}) gives the values of special coordinates at infinity, these
are
\begin{equation}
X^0_\infty={1\over h_0}, \quad X^1_\infty={1\over h_1}, \quad
X^2_\infty={1\over h_2}
\end{equation}
and it can be easily seen that the mass of the black hole saturates the 
BPS bound
\begin{equation}
M_{ADM}= {\pi\over 4G_N} Z_\infty, 
\end{equation}
where $Z_\infty$ is the value of the central charge at infinity,
$Z_\infty={q_0\over h_0}+{q_1\over h_1}+{q_2\over h_2}$.

Moreover, from (\ref{pain}) one could easily see that irrespective of the
values of the moduli at infinity, the values, near the horizon 
$(r\rightarrow 0)$, $(S, T, U)$ are always given by
\begin{equation}
S_{hor}=\Big({q_1q_2\over q_0^{2}}\Big)^{{1\over3}},\qquad
T_{hor}=\Big({q_0q_2\over q_1^{2}}\Big)^{{1\over3}},\qquad
U_{hor}=\Big({q_0q_1\over q_2^{2}}\Big)^{{1\over3}}.
\label{cons}
\end{equation}
These solutions can also be derived from the stabilisation equations 
$ZX_I=q_I$
\cite{chaka}. The central charge near the horizon is thus given by 
$Z_{hor}=3{(q_0q_1q_2)^{1\over3}}$.  

Moreover, the Bekenstein-Hawking entropy ${\bf S}_{BH}$ which is defined in 
terms of the 3-area of the regular $r=0$ horizon ${\bf A}$, is given by
\begin{equation}
{\bf S}_{BH}={{\bf A}\over 4G_N};\qquad {\bf A}=
2\pi^2\Big({Z\over 3}\Big)^{3\over2}=2\pi^2\sqrt{q_0q_1q_2}.
\label{shadi}
\end{equation}
Finally, the double extreme limit, where the values of the moduli are
constant everywhere and are equal to their values at the horizon given by
(\ref{cons}), corresponds to an extreme solution with the choice
\begin{equation}
h_0=\Big({q_0^{2}\over q_1q_2 }\Big)^{{1\over3}},\qquad
h_1=\Big({q_1^{2}\over q_0q_2 }\Big)^{{1\over3}},\qquad
h_1=\Big({q_2^{2}\over q_0q_1}\Big)^{{1\over3}}
\label{co}
\end{equation}
and thus the metric for the double extreme solution is given by
\begin{equation}
ds^2=-\Big(1+{(q_0q_1q_2)^{1\over3}\over r^2}\Big)^{-2}dt^2+
\Big(1+{(q_0q_1q_2)^{1\over3}\over r^2}\Big)(dr^2+r^2d\Omega_3^2).
\end{equation}

For the choice, $h_1=h_2=h_3=1$, one obtains the solution
\begin{equation}
ds^2=-\Big((1+{q_0\over r^2})(1+{q_1\over r^2})(1+{q_2\over r^2})\Big)^
{-2\over 3}dt^2+\Big((1+{q_0\over r^2})(1+{q_1\over r^2})(1+{q_2\over r^2})
\Big)^{1\over3}
\end{equation}
which is equivalent to the five-dimensional black hole constructed in
\cite{Tsey}. Clearly the entropy is still given by (\ref{shadi}) since it is
independent of the values of $h's$.

\section{Conclusions}
In this paper, we have constructed general 
BPS black hole solutions for $N=2$, $d=5$
supergravity coupled to an arbitrary number of vector multiplets. 
We have demonstrated that these solutions 
admit supersymmetry by solving for the
supersymmetry transformation laws for the gauginos and the
gravitino, making use of the underlying very special geometry
structure which governs the couplings in these theories. Using our explicit
solution, we have verified that the horizon acts as an attractor on which the
scalar fields take constant values which extremise the central charge.
In other words, the equations defining the special
coordinates in our solutions reduce, near the horizon, to the stabilisation
equations obtained from the extremisation of the central charge 
of the underlying $N=2$, $d=5$ supersymmetry algebra. 
The fixed values are independent of the initial configuration at infinity.
The dependence of the entropy on the central charge is also derived.
As an example we considered
the $STU$ model in some detail and derived various physical quantities
characterising its black hole solution.

Recently, it became obvious that BPS black holes play a 
major role in the analysis of phase transitions 
among $N=2$ strings vacua. Generically, these topological phase transitions 
occur at points in the moduli spaces where non-perturbative BPS states
become massless.  For example, in $N=2$ type IIB string vacua
on Calabi-Yau three-folds, electrically or magnetically charged black
holes become massless at the conifold points in the Calabi-Yau moduli
spaces, where certain homology cycles of the Calabi-Yau spaces shrink
to zero size \cite{conifold}. 
Non-trivial black hole solutions provide the framework in which
one can find those points in space-time where the 
internal Calabi-Yau periods shrink to zero size and where the transitions 
due to massless black holes take place \cite{latest}.

In Calabi-Yau compactification of type IIA, due to the presence of theta
angles, the vector moduli space can be described in terms of a complexified
K\"ahler cone \cite{transitions}. As such, as one approaches the 
boundary of the K\"ahler cone, one can go smoothly past the 
singularity to another phase of the
conformal field theory \cite{transitions}. 
The new phase might correspond to a new K\"ahler cone of
a different Calabi-Yau, thus signalling a topology changing process, or it
might be related to a non-geometrical phase, $i.e$ an abstract formulation of
the conformal field theory, such as Landau-Ginzburg model.

In five dimensional $N=2$ supergravity theories 
corresponding to M-theory compactification on a Calabi-Yau threefold, 
there are no axionic fields and the moduli, the sizes of 2-cycles in the
Calabi-Yau, take values in a real K\"ahler cone. In contrast to the 
four-dimensional case, to go from one phase to another one has to go through
the singularity. As a result, in five dimensions
one gets sharp phase transition \cite{witten}. One example of such transitions
is the "flop transition", where one of the moduli approaches 
zero and subsequently 
blown up, this corresponds to going into a new K\"ahler cone of a birationally
equivalent Calabi-Yau with different intersection numbers. 
In analysing these flop transitions, BPS black holes
which become massless at the phase transition point play a significant role 
\cite{witten}. 
Recently, the BPS central charge $Z$, 
as well as the magnetic charge 
defining the tension of the five-dimensional string have been employed in  
probing the flop phase transitions in five dimensions \cite{chaka}. 
It would be of interest to study these transitions 
using our explicit black hole solutions along the lines of \cite{latest}. This
is currently under investigation.

{\large \bf Acknowledgements}
I would like to thank K. Behrndt, D. L\"ust and T. Mohaupt for useful
conversations. This work is supported by DFG and in part by DESY-Zeuthen.

\end{document}